\documentclass{pasj00}
\draft
\begin{document}
\Received{2012/9/4}
\Accepted{2012/9/28}
\title{Measurements of Antenna Surface for a Millimeter-Wave Space Radio Telescope II;\\
Metal Mesh Surface for Large Deployable Reflector}
\SetRunningHead{\textsc{Kamegai} \rm et al. }{Antenna Surface}

\author{
	Kazuhisa \textsc{Kamegai}\altaffilmark{1,2}
	\and
	Masato \textsc{Tsuboi}\altaffilmark{1}
}
\altaffiltext{1}{Institute of Space and Astronautical Science (ISAS), Japan Aerospace Exploration Agency (JAXA), 3-1-1 Yoshinodai, Chuo-ku, Sagamihara, Kanagawa 252-5210, Japan}
\altaffiltext{2}{Department of Industrial Administration, Faculty of Science and Technology, Tokyo University of Science, Yamazaki 2641, Noda, Chiba 
278-8510, Japan}

\maketitle

\begin{abstract}
Large deployable antennas with a mesh surface woven by fine metal wires are an important technology for communications satellites and space radio telescopes. However, it is difficult to make metal mesh surfaces with sufficient radio-frequency (RF) performance for frequencies higher than millimeter waves. 
 In this paper, we present the RF performance of metal mesh surfaces at 43~GHz. 
For this purpose, we developed an apparatus to measure the reflection coefficient, transmission coefficient, and radiative coefficient of the mesh surface.
The reflection coefficient increases as a function of metal mesh surface tension, whereas the radiative coefficient decreases. 
The anisotropic aspects of the reflection coefficient and the radiative coefficient are also clearly seen. They depend on the front and back sides of the metal mesh surface and the rotation angle.
The transmission coefficient was measured to be almost constant. 
The measured radiative coefficients and transmission coefficients would cause significant degradation of the system noise temperature.
In addition, we carried out an astronomical observation of a well-known SiO maser source, R Cas, by using a metal mesh mirror on the NRO 45-m radio telescope Coud\'e system. 
The metal mesh mirror considerably increases the system noise temperature and slightly decreases the peak antenna temperature. These results are consistent with laboratory measurements.
\end{abstract}
 
\section{Introduction} \label{sec:intro}
The history of radio telescope development follows the quest for higher angular resolution. High angular resolution, no worse than milliarcseconds, is essential to observe the innermost regions of active galactic nuclei (AGN) and astronomical masers. The angular resolution of a radio interferometer is given by $\lambda/D$, where $\lambda$ is wavelength and $D$ is the baseline length or antenna spacing. Ground-based interferometry, including very long baseline interferometry (VLBI), has an upper limit of angular resolution because $D$ cannot exceed Earth's diameter, approximately 13,000~km. Space VLBI was formulated to extend this limit. In this method, a radio interferometer connects orbital radio telescopes with ground-based ones. Since the baseline length of space VLBI can extend beyond the diameter of Earth, the angular resolution of space VLBI can exceed ground-based limits. This concept was shown to be feasible by the demonstration using NASA's Tracking and Data Relay Satellite (TDRS) and ground-based telescopes (\cite{Levy1986})

The first astronomical space VLBI observations were conducted by the VLBI Space Observatory Programme (VSOP; \cite{Hirabayashi1998}) with the HALCA satellite (\cite{Hirabayashi2000}). Recently Russian spacecraft Spektr-R was launched successfully and RadioAstron (\cite{Kardashev1997}) was started as the second astronomical space VLBI project.
ASTRO-G was developed by ISAS/JAXA as a successor to these radio telescope satellites  (\cite{Tsuboi2008}, \cite{Saito2009}, \cite{Saito2010}) 
with observation frequency bands at $8.0$--$8.8$, $20.6$--$22.6$, and $41$--$45$~GHz.
The main reflector on the radio telescope antenna is a 9.2~m deployable offset-Cassegrain large deployable reflector (LDR; \cite{Higuchi2009}). The LDR is composed of seven 4~m hexagonal modules, which have a radial-rib/hoop-cable back structure and metal mesh surface, allowing high surface accuracy with a small payload (\cite{Higuchi2009}). 
Because metal mesh surfaces perform well up to centimeter-wave frequencies (e.g., \cite{Imbriale1991}), 
they are commonly used as antenna surfaces, for example in NASA's  TDRS (\cite{Tankersley1977}), JAXA's Engineering Test Satellite-VIII (ETS-VIII; \cite{Meguro2003}), and the HALCA satellite (\cite{Takano2004}, \cite{Hanayama2004}). The radio-frequency (RF) performance of metal mesh surfaces has been measured up to 22~GHz (e.g., \cite{Hanayama1993}). They have not been reported over 40~GHz, however, despite the importance of precise measurements of RF performance for the development of millimeter-wave space radio telescope antennas. This work was performed during antenna development of the ASTRO-G program which was unfortunately canceled. However, the knowledge gained is indispensable for the development of future millimeter-wave space radio telescope antennas. In this paper, we present the measurement method, its results, and points for consideration when constructing antennas.

\section{Mesh Structure and Specimens}\label{sec:mesh}
\begin{figure}
	\begin{center}
	\FigureFile(80mm,60mm){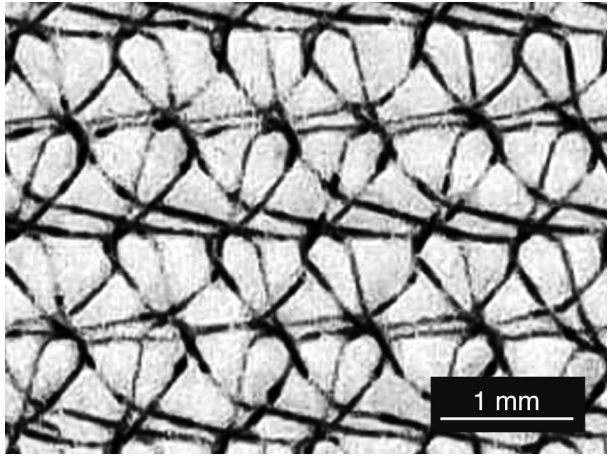}
	\end{center}
	\caption{Photograph of the metal mesh surface. The fabric is made of gold-plated molybdenum wires of 30~$\mu {\rm m}$ diameter.
	}
	\label{fig:photo_mesh}
\end{figure}

  Figure \ref{fig:photo_mesh} shows the appearance of the metal mesh surface for ASTRO-G's LDR. It is woven with gold-plated molybdenum wires of 30~$\mu {\rm m}$ in diameter. The metal mesh fabric employs a single satin stitch weaving pattern, because that pattern has relatively good isotropic mechanical retractility and electric performance at 20~GHz (\cite {Miura2003}). 
This is a type of warp-knitted fabric where wires flow in the vertical direction and jump across two neighboring stitches at every horizontal row, giving the metal mesh surface front and back sides.
The knitting density is 28 stitches per inch, which corresponds to a $\sim 0.9$~mm stitch size. The ratios of the stitch size to the wavelength at 8, 22, and 43~GHz are $\lambda/42$, $\lambda/15$, and $\lambda/8$, respectively. 

We use four approximately 200~mm $\times$ 180~mm specimens of metal mesh material for our measurements. Tensions of 100, 350, 500, and 700~gf/m are placed isotropically and homogeneously on the mesh of each specimen. 
The incident angle of the radio frequency to the specimen is $45^{\circ}$, due to conditions of the measurement apparatus (see section \ref{sec:method}). 
Except in the case of a metal plate surface, the RF performance of the metal mesh surface depends on which side is used, and the rotation angle of the specimen on which the incident wave reflects. 
For this reason, we make measurements on both sides and at four rotation angles, a total of eight conditions for each tension setting.

\begin{figure}
	\begin{center}
	\FigureFile(80mm,50mm){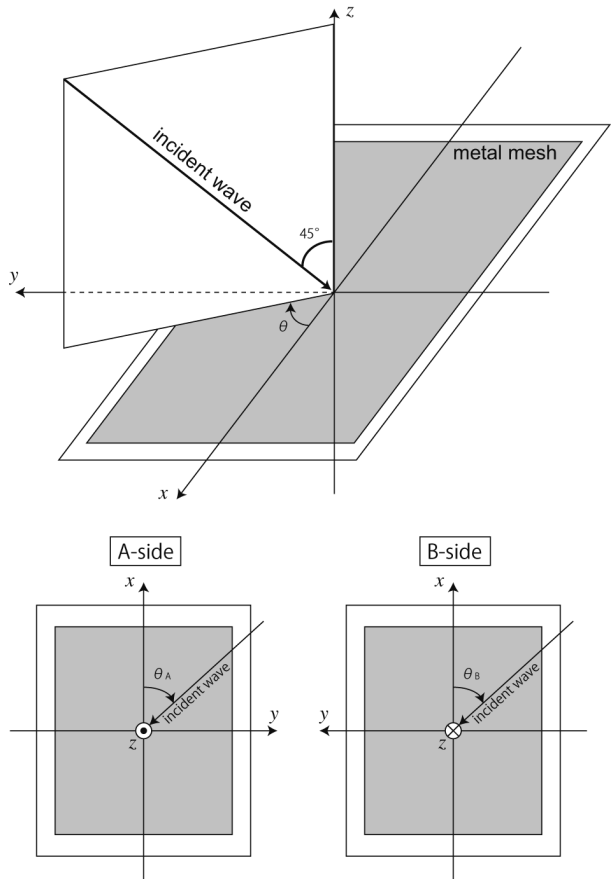}
	\end{center}
	\caption{The coordinate system for the metal mesh surface specimens is set as  in the upper figure. The incident wave reflects at $45^{\circ}$ on the specimen surface due to conditions of the measurement apparatus (see section \ref{sec:method}). The rotation angles on each side ($\theta_{\rm A}$ and $\theta_{\rm B}$) are defined as in the lower figures.}\label{fig:coordinates}
\end{figure}

The side of the specimen on which the incident wave gives the largest reflection coefficient is defined as the A-side, and the opposite side is the B-side. The coordinate system is set on the A-side as shown in Fig. \ref{fig:coordinates}. The $x$-axis is defined as the incident wave direction with the largest reflection coefficient. The $y$-axis is set perpendicular to the $x$-axis. As a result, the $x$- and $y$-axes are along the long and short sides of the specimen, respectively. Then the $z$-axis is set to be perpendicular to the $x$--$y$ plane, and is positive on the A-side.
Rotation of the specimen is defined with angles in the direction clockwise from the $x$-axis (Fig. \ref{fig:coordinates}). Rotation angles on the A-side and B-side are described as $\theta_{\rm A}$ and $\theta_{\rm B}$, respectively.

\section{Measurement Method of RF performance} \label{sec:method}
We measured the Q band ($41$--$45$~GHz) RF performance for four metal mesh surface tensions at two side and four rotation angles (32 conditions total). As mentioned previously, onboard radio telescopes have much stricter noise requirements than do communications antennas. However, no measurement apparatus existed for such low reflection loss at millimeter-wave frequencies. We therefore developed a measurement apparatus, described in a previous paper (\cite {Kamegai2011}). Measurements in this paper use right-handed circular polarization (RCP). 

\begin{figure}
	\begin{center}
	\FigureFile(80mm,59mm){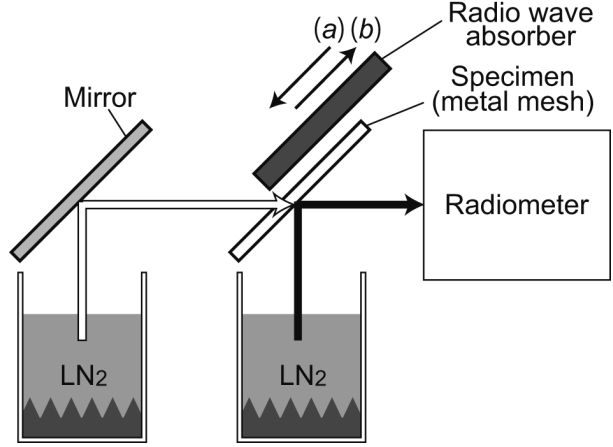}
	\end{center}
	\caption{Configuration of the apparatus used in our measurements. The apparatus consists of a radiometer, a specimen holder, and terminating equipment for transmission. Details of the radiometer configuration are described in \cite{Kamegai2011}. The black and the white arrow represent waves incident to the radiometer by reflection and by transmission at the specimen, respectively. The measurement was carried out under the following two conditions for transmission: ({\it a}) When a radio absorber is inserted behind the specimen, the transmission is terminated with the room temperature load (see also the text and equation (\ref{eq4})). ({\it b}) When the radio absorber is removed, the transmission is terminated with the liquid nitrogen temperature load (see also the text and equation (\ref{eq5})).}\label{fig:measurements}
\end{figure}

Here we introduce the measurement method for reflection loss. Reflection loss consists of two distinct terms, the metal mesh's radiative coefficient $\alpha$ and its transmission coefficient $\beta$. From conservation of energy, the relation among these and the reflection coefficient $\gamma$ is simply given by 
\begin{equation}
\label{eq1}
\alpha+\beta+\gamma=1.
\end{equation}
 When the radiometer's input flange is terminated with a room temperature load, the output $P_{\rm hot, 0}$ is given by
 \begin{equation}
\label{eq2}
P_{\rm hot, 0} = G_rkB(T_{\rm RX} + T_{\rm amb}), 
\end{equation}
where $T_{\rm RX}$ and $T_{\rm amb}$ are the noise temperature of the radiometer receiver and room temperature, respectively. $B$ and $G_r$ are the frequency bandwidth and the radiometer receiver gain, respectively. 
The $T_{\rm RX}$ is kept as low as $\sim 30$~K throughout all measurements.
The output when terminated with a liquid nitrogen temperature load $P_{\rm cold, 0}$ is given by 
\begin{equation}
\label{eq3}
P_{\rm cold, 0} = G_rkB(T_{\rm RX} + T_{\rm LN_{2}}), $$
\end{equation}
where $T_{\rm LN_{2}}$ is liquid nitrogen temperature.
Figure \ref{fig:measurements} shows how the metal mesh specimen is inserted. 
When the transmission is terminated with a room temperature load, the output from the radiometer $P_{\rm hot, m}$ is given by
\begin{equation}
\label{eq4}
P_{\rm hot, m} = G_rkB(T_{\rm RX} + \alpha T_{\rm amb} + \beta T_{\rm amb} + \gamma T_{\rm LN_{2}}).
\end{equation}
When the transmission is terminated with a liquid nitrogen temperature load, the output from the radiometer $P_{\rm cold, m}$ is given by
\begin{equation}
\label{eq5}
P_{\rm cold, m} = G_rkB(T_{\rm RX} + \alpha T_{\rm amb} + \beta T_{\rm LN_{2}} + \gamma T_{\rm LN_{2}}).
\end{equation}

The combination of these formulas leads to
\begin{equation}
\label{eq6}
\alpha = \frac{P_{\rm cold, m}- P_{\rm cold, 0}}{P_{\rm hot, 0} -P_{\rm cold, 0}},
\end{equation}

\begin{equation}
\label{eq7}
\beta = \frac{P_{\rm hot, m}- P_{\rm cold, m}}{P_{\rm hot, 0} -P_{\rm cold, 0}}, 
\end{equation}

and

\begin{equation}
\label{eq8}
\gamma = \frac{P_{\rm hot, 0}- P_{\rm hot, m}}{P_{\rm hot, 0} -P_{\rm cold, 0}}.
\end{equation}
We evaluated the coefficients of the mesh mirror, $\alpha$, $\beta$, and $\gamma$, using these formulas.  

Measurements were carried out for the specimen rotation angles $\theta_{\rm A} =0^{\circ}$, $90^{\circ}$, $180^{\circ}$, $270^{\circ}$ and $\theta_{\rm B} = 0^{\circ}$, $90^{\circ}$, $180^{\circ}$, $270^{\circ}$. 

\section{Results of RF Performances} \label{sec:result}
\begin{figure}
	\begin{center}
	\FigureFile(80mm,130mm){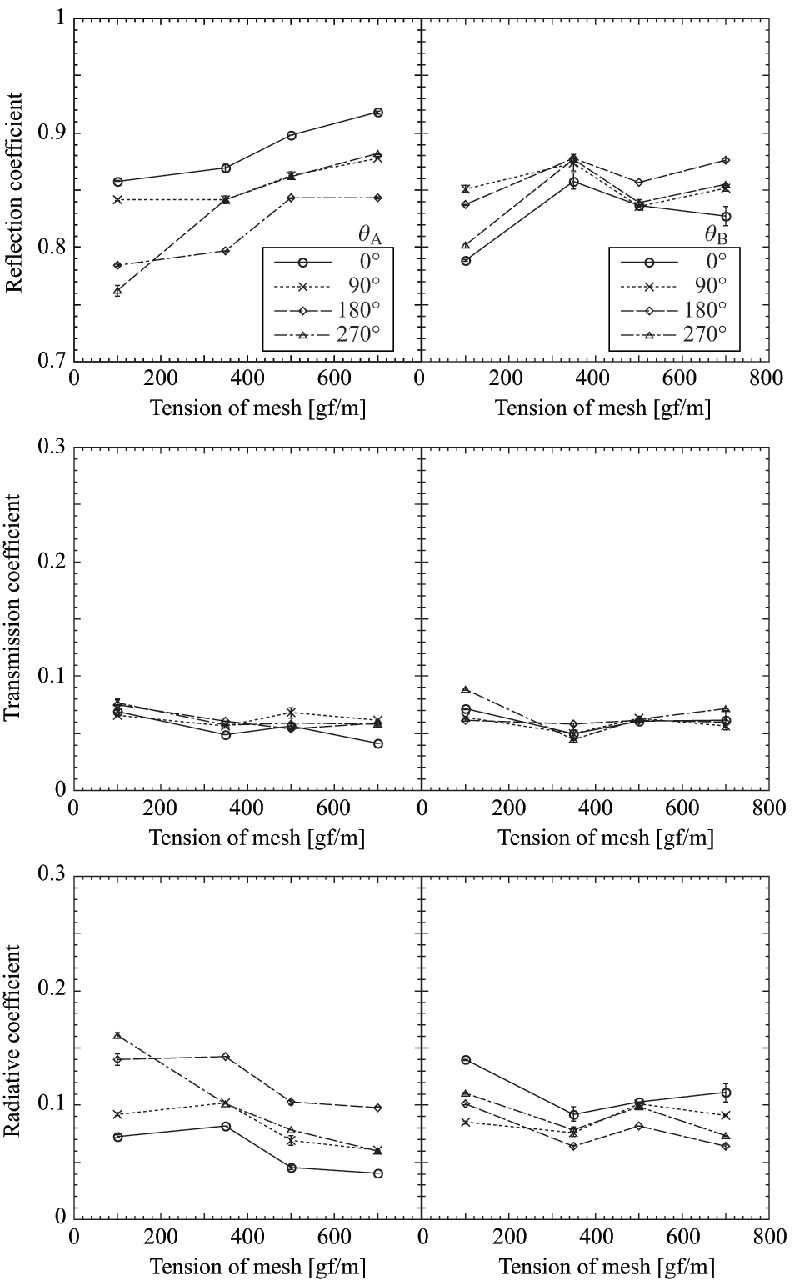}
	\end{center}
	\caption{Results of measurements of reflection coefficient ({\it upper panels}), transmission coefficient ({\it middle panels}), and radiative coefficient ({\it bottom panels}). The left and right panels indicate the results of the A-side and B-side of the specimen, respectively.}
	\label{fig:results}
\end{figure}

Table \ref{table:results} and Figure \ref{fig:results} show reflection coefficient $\gamma$ (upper panels), transmission coefficient $\beta$ (middle panels), and radiative coefficient $\alpha$ (lower panels) as a function of the metal mesh surface tension. Error was estimated as the standard deviation of five measurements for each condition. In the following subsections, we describe the results in detail.

\begin{table*}[htdp]
\caption{Measurement results}
\begin{center}
\tiny
\begin{tabular}{cccccccccccc}
\hline
\hline
& & \multicolumn{4}{c}{Reflection Coefficient} & & & \multicolumn{4}{c}{Reflection Coefficient} \\
\cline{3-6} \cline{9-12}
Side & $\theta_{\rm A}$ & \multicolumn{4}{c}{tension [gf/m]} & Side & $\theta_{\rm B}$ & \multicolumn{4}{c}{tension [gf/m]} \\
& [deg] & 100 & 350 & 500 & 700 & & [deg] & 100 & 350 & 500 & 700 \\
\hline
A & 0 & 0.8580(7) & 0.8696(39) & 0.8981(1) & 0.9179(10) & B & 0 & 0.7886(9) & 0.8578(73) & 0.8369(1) & 0.8275(83) \\

A & 90 & 0.8417(9) & 0.8417(30) & 0.8629(29) & 0.8781(4) & B & 90 & 0.8512(30) & 0.8740(69) & 0.8361(7) & 0.8520(26)\\

A & 180 & 0.7844(8) & 0.7966(2) & 0.8431(3) & 0.8429(8) & B & 180 & 0.8373(4) & 0.8779(2) & 0.8569(2) & 0.8762(18)\\

A & 270 & 0.7621(45) & 0.8415(13) & 0.8620(10) & 0.8818(4) & B & 270 & 0.8016(23) & 0.8766(25) & 0.8389(28) & 0.8551(6)\\

\hline
& & \multicolumn{4}{c}{Transmission Coefficient} & & & \multicolumn{4}{c}{Transmission Coefficient} \\
\cline{3-6} \cline{9-12}
Side & $\theta_{\rm A}$ & \multicolumn{4}{c}{tension [gf/m]} & Side & $\theta_{\rm B}$ & \multicolumn{4}{c}{tension [gf/m]} \\
& [deg] & 100 & 350 & 500 & 700 & & [deg] & 100 & 350 & 500 & 700 \\
\hline

A & 0 & 0.0694(13) & 0.0488(2) & 0.0561(18) & 0.0413(0) & B & 0 & 0.0716(4) & 0.0501(63) & 0.0606(4) & 0.0617(82)\\

A & 90 & 0.0660(9) & 0.0564(1) & 0.0679(44) & 0.0613(3) & B & 90 & 0.0640(11) & 0.0497(35) & 0.0632(18) & 0.0567(7)\\

A & 180 & 0.0754(50) & 0.0609(2) & 0.0541(21) & 0.0589(2) & B & 180 & 
0.0614(7) & 0.0581(12) & 0.0612(2) & 0.0596(16)\\

A & 270 & 0.0771(23) & 0.0570(4) & 0.0594(2) & 0.0583(6) & B & 270 & 0.0883(8) & 0.0451(27) & 0.0622(23) & 0.0717(4)\\

\hline

& & \multicolumn{4}{c}{Radiative Coefficient} & & & \multicolumn{4}{c}{Radiative Coefficient} \\
\cline{3-6} \cline{9-12}
Side & $\theta_{\rm A}$ & \multicolumn{4}{c}{tension [gf/m]} & Side & $\theta_{\rm B}$ & \multicolumn{4}{c}{tension [gf/m]} \\
& [deg] & 100 & 350 & 500 & 700 & & [deg] & 100 & 350 & 500 & 700 \\
\hline

A & 0 & 0.0726(13) & 0.0816(2) & 0.0458(18) & 0.0408(0) & B & 0 & 
0.1398(4) & 0.0921(63) & 0.1025(4) & 0.1109(82)\\

A & 90 & 0.0923(9) & 0.1019(1) & 0.0693(44) & 0.0606(3) & B & 90 & 
0.0849(11) & 0.0763(35) & 0.1008(18) & 0.0912(7)\\

A & 180 & 0.1402(50) & 0.1425(2) & 0.1028(21) & 0.0982(2) & B & 180 & 0.1013(7) & 0.0640(12) & 0.0819(2) & 0.0642(16)\\

A & 270 & 0.1608(23) & 0.1015(4) & 0.0786(2) & 0.0600(6) & B & 270 & 0.1101(8) & 0.0783(27) & 0.0989(23) & 0.0732(4)\\

\hline

\end{tabular}

\renewcommand{\thefootnote}{}
\footnotemark{Numbers in parentheses represent measurement error in units of the last significant digits.}
\end{center}
\label{table:results}
\end{table*}%

\subsection{Reflection Coefficient}\label{subsec:ref}
The reflection coefficient $\gamma$ gradually increases as a function of the metal mesh tension. For example, when $\theta_{\rm A}$ is $0^{\circ}$, $\gamma$ increases from 0.86 at 100~gf/m to 0.92 at 700~gf/m. 
This trend is pronounced on the A-side, whereas reflection coefficients for the B-side flatten for tensions greater than 350~gf/m.
\cite {Hanayama1993} suggested that electrical currents induced by incident RF depend on contact resistance between the mesh wires, which may explain this trend. Higher tension causes tighter contact between the wires, lowering contact resistance. Currents therefore flow more easily at higher tension, raising the reflection coefficient. 

Rotation angle dependencies are also found. When the incident wave comes from the positive direction of the $z$-axis, the reflection coefficient is maximized at $\theta_{\rm A} = 0^{\circ}$ and minimized at $\theta_{\rm A} = 180^{\circ}$. The converse occurs for incident waves from the negative direction of the $z$-axis; the minimum value is at $\theta_{\rm A} = 0^{\circ}$ and the maximum value is at $\theta_{\rm A} = 180^{\circ}$.
As for $\theta_{\rm A} = 90^{\circ}$ and $270^{\circ}$, the reflection coefficients were intermediate values between those of $\theta_{\rm A} = 0^{\circ}$ and $180^{\circ}$. Values agreed within 0.01, except for the case of 100~gf/m tension. These differences between the rotation angles of the specimen seem to be attributed to the anisotropy of the mesh knitting pattern, as mentioned in \cite {Hanayama1993}. In addition, the overlapping pattern of wires may contribute to the fact that the largest differences are measured between $0^{\circ}$ and $180^{\circ}$, and that the relationship between the two angles is reversed between sides of the specimens.

\subsection{Transmission Coefficient}\label{subsec:trans}
The transmission coefficient $\beta$ is measured as $0.041$--$0.088$ for all conditions. No dependence on tension, side, or specimen rotation angle was evident, suggesting that the transmission coefficient is determined only by the mechanical structure of the metal mesh, such as pattern and density of the fabric. 

\cite {Hanayama1993} reported transmission loss in the 22~GHz band toward a metal mesh surface of doubly inlaid knitting for the condition of six incident angles ($0^{\circ}$--$90^{\circ}$ with $15^{\circ}$ step), two rotation angles of the mesh ($0^{\circ}$ and $90^{\circ}$), and two polarization angles of the incident wave (parallel and perpendicular to the plane of incidence). 
In the case of the incident angle of $45^{\circ}$, their results showed that difference in transmission loss between rotation angles of $0^{\circ}$ and $90^{\circ}$ was measured to be 0.12 when the direction of electric field of incident wave was parallel to the plane of incidence and 0.03 when those were perpendicular to each other. Including differences between the polarization angles, the difference between maximum and minimum values exceeded 0.16. 

\cite {Miura2003} also reported transmission losses of metal meshes with single atlas, back half tricot, and single satin stiches at 20~GHz. Measurements were for specimen rotation angles of $0^{\circ}$, $45^{\circ}$, and $90^{\circ}$ with respect to the direction of electric field of the measured linearly polarized wave. Their results showed differences between the maximum and the minimum values for the above three knitting patterns to be 0.21, 0.020, and 0.006, respectively, suggesting single satin as the best material for an LDR antenna. 

Our results for single satin in the Q band with circular polarization showed a transmission coefficient at 500~gf/m tension in the range $0.054$--$0.068$ (Table \ref{table:results}), a difference of 0.014. This indicates that transmission anisotropy is small under our measurement conditions as well.

\subsection{Radiative Coefficient}\label{subsec:rad}
Contrary to the reflection coefficient, the radiative coefficient generally decreases as a function of the metal mesh tension. 
The rotation angle dependency trend is also contrary to that of the reflection coefficient. On the A-side, a $\theta_{\rm A} = 180^{\circ}$ rotation angle has the largest radiative coefficients for all tension values except for 100~gf/m. Rotation angles $\theta_{\rm A} = 90^{\circ}$ and $270^{\circ}$ have similar radiative coefficients, and the second largest values among each tension. Minimum values are measured at $\theta_{\rm A} = 90^{\circ}$.

Among all measurements, the maximum value was as high as 0.16 at 100~gf/m tension and $\theta_{\rm A} = 270^{\circ}$, whereas the minimum value was 0.04 at 700~gf/m tension and $\theta_{\rm A} = 0^{\circ}$. Since the measured transmission coefficients indicate similar values, the trend can be interpreted as being a result of conservation of energy.

\section{Increase of the System Noise Temperature by Losses}\label{sec:tsys}
 When a space radio telescope observes blank sky, the output power of the telescope is given by
\begin{equation}
\label{eq9}
P_{\nu}(sky)=Gk(T_{RX}+\alpha T_{\rm amb}+\beta T_{\rm back}+\gamma T_{\rm CBR}),
\end{equation}
where $T_{\rm RX}$, $T_{\rm amb}$, and $T_{\rm back}$ are the receiver noise temperature, the physical temperature of the main reflector, and the radiation temperature from the back side of the antenna, respectively. $ T_{\rm CBR}$ is the contribution of the cosmic microwave background, $2.7$~K. $G$ is the receiver gain. Here we ignore the negligible contribution of the sub-reflector (Kamega et al. 2011). The total system noise temperature of the space telescope $T_{\rm sys}$ is given by 
\begin{equation}
\label{eq10}
T_{\rm sys}=\frac{P_\nu(\rm{sky})}{\gamma Gk}
			 =\frac{T_{RX}+\alpha T_{\rm amb}+\beta T_{\rm back}+\gamma T_{\rm CBR}}{\gamma}.
\end{equation}

In the case of the ASTRO-G, the originally required $T_{\rm sys}$ at the Q band was less than $T_{\rm sys}\lesssim72$~K. The first term of equation (\ref{eq10}) is the contribution of receiver noise. Since the measured receiver noise temperature is less than $T_{\rm RX}\lesssim60$~K, its contribution is estimated to be $T_{\rm RX}/\gamma=65$~K when we use the A-side at the rotation angle $\theta_{\rm A} = 0^{\circ}$ and 700~gf/m tension. If we use the lower tension of the metal mesh surface or other rotation angle, the contribution is dramatically increased to $T_{\rm RX}/\gamma \lesssim 79$~K. 
Control of these contributions is important to realize the high reflection constant required by antenna surfaces. The last term is the contribution of the cosmic microwave background. 

If the contribution of the main reflector is less than 3~K, $T_{\rm sys}$ meets the original requirement of the ASTRO-G. 
However, noise in the transmission coefficient and radiative coefficient (formulated as the second and third term in equation (\ref{eq10}), respectively) could cause a significant increase of the system noise temperature $T_{\rm sys}$, due to the radiative coefficient $\alpha$, as follows. 
\begin{equation}
\label{eq11}
\Delta T_{\rm sys,\alpha} = \frac{\alpha T_{\rm amb}}{\gamma}.
\end{equation}
The physical temperature of the ASTRO-G LDR in orbit is expected to reach up to $T_{\rm amb}=490$~K, which makes $\Delta T_{\rm sys, \alpha}$ maximum. When we use the A-side at the rotation angle $\theta_{\rm A}=0^{\circ}$ and 700~gf/m tension, $\Delta T_{\rm sys, \alpha}$ is estimated to be 22~K. When we use the A-side at the  rotation angle of $\theta_{\rm B}=270^{\circ}$ and 100~gf/m tension, it is evaluated to be as high as 103~K.

On the other hand, additional noise by transmission depends on the radiation temperature from the backside of the antenna, as follows. 
\begin{equation}
\label{eq12}
\Delta T_{\rm sys, \beta}=\frac{\beta T_{\rm back}}{\gamma}.
\end{equation}
Assuming a backing Earth, $T_{\rm back}$ is 300~K. Then $\Delta T_{\rm sys, \beta}$ is calculated to be $14$--$33$~K. 
On the other hand, with blank sky behind the antenna the cosmic microwave background can be used as $T_{\rm back}$. Then the $\Delta T_{\rm sys, \beta}$ is estimated to be $0.12$--$0.30$~K, which is almost negligible. The increase in $T_{\rm sys}$ would therefore exceed the ASTRO-G requirement only in the case where Earth is behind the antenna.

The estimated total system noise temperature is therefore
\begin{equation}
\label{eq13}
T_{\rm sys} = \frac{T_{RX}}{\gamma}+\Delta T_{\rm sys,\alpha} + \Delta T_{\rm sys,\beta}+T_{\rm CBR}=90-185~{\rm K}
\end{equation}
when Earth is not behind the antenna, 
and $T_{\rm sys} =103-215~{\rm K}$ when it is. In both cases, the minimum value was measured on the A-side with $\theta_{\rm A}=0^{^\circ}$ and 700~gf/m tension, whereas the maximum value was measured on the A-side with $\theta_{\rm A}=270^{^\circ}$ and 100~gf/m tension.
Even the minimum value of the estimated $T_{\rm sys}$ is much higher than the ASTRO-G requirement and could significantly degrade sensitivity, though astronomical observations would be not impossible. 
To suppress the increasing of $T_{\rm sys}$ when the LDR is put into production, our results suggest that tension of the metal mesh surface should be kept as tight as possible, at least in the range of our measurements. The angle of the metal mesh equipped on the antenna surface should also be considered.

\section{RF Performance Test Observation with the Nobeyama 45-m Telescope} \label{sec:nro}
\begin{figure}
	\begin{center}
	\FigureFile(80mm,111mm){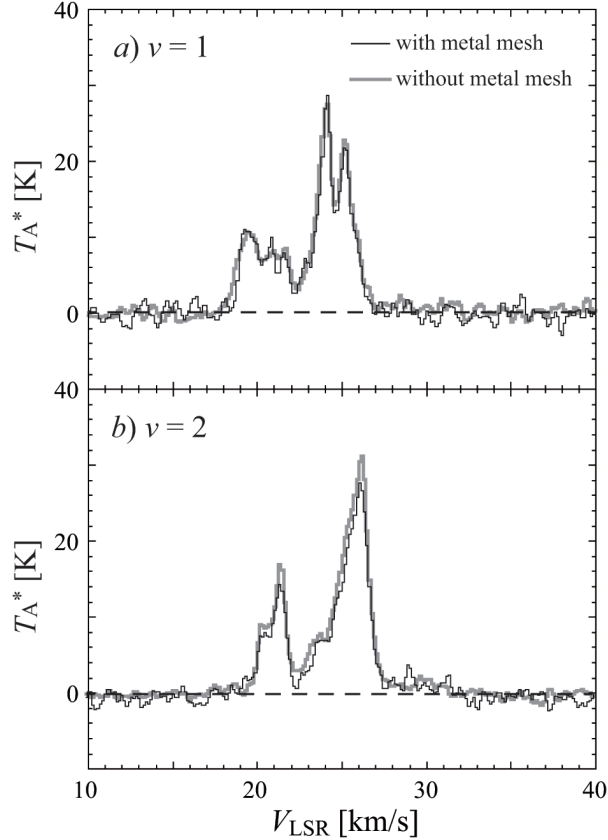}
	\end{center}
	\caption{Spectra of SiO $J=1$--$0$, $v=2$ maser lines at 43~GHz toward R-Cas with the Nobeyama 45-m telescope. Thick gray line and thin black line indicate the results with and without the metal mesh attached to the beam guide system of the telescope.}\label{fig:nro}
\end{figure}

We verified the RF performance of the metal mesh using the 45-m radio telescope of the Nobeyama Radio Observatory (NRO) \footnote{The Nobeyama Radio Observatory is a branch of National Astronomical Observatory, National Institutes of Natural Sciences, Japan} to confirm whether it meets the requirement as a radio telescope antenna surface. A metal mesh flat mirror was set on the metal solid mirror of the telescope's beam guide system. We use the A-side at an rotation angle of $\theta_{\rm A}=0^{\circ}$ and 100~gf/m tension.
We used the telescope to observe the SiO $J=1$--$0$, $v=1$ (43.122~GHz), and $v=2$ (42.821~GHz) maser emission lines for a well-known SiO maser source, R-Cas. 
After removing the metal mesh mirror, we immediately observed the same object. Figure \ref{fig:nro} shows both spectra. The system noise temperatures with and without the metal mesh mirror were 277~K and 209~K, respectively. The peak antenna temperature of the SiO maser lines with the metal mesh mirror was 0\% at $v=1$ and 10\% at $v=2$ less than that without it. 

The feed efficiency of the telescope is given by 
\begin{equation}
\label{eq14}
\eta =1-\zeta-\kappa,
\end{equation}
where $\zeta$ is the sum of the fractions of Ohmic loss of the mesh surface $\alpha$ and the fraction of the transmitted radiation terminated by the telescope inside $\beta$ (i.e., $\zeta=\alpha+\beta$), and $\kappa$ is the fraction of the scattered radiation terminated by the sky. $\kappa$ is not measured in the previous sections. When the ground-based telescope observes blank sky, the output power of the telescope is 
\begin{equation}
\label{eq15}
P_\nu(sky)=Gk[T_{\rm RX}+\zeta T_{\rm amb}+\eta(1-e^{-\tau})T_{\rm atm}],
\end{equation}
where $T_{\rm amb}$ is the physical temperature of the telescope, $T_{\rm atm}$ is the physical temperature of the atmosphere, and $\tau$ is the optical thickness of the atmosphere along the telescope beam. When the telescope observes an object, the output power is 
\begin{equation}
\label{eq16}
P_\nu(source)=GkT_{\rm A}+P_\nu(sky).
\end{equation}
where $T_{\rm A}$ is the antenna temperature measured at the output of the receiver.
When the receiver input feed is terminated by the ambient temperature absorber, the output power is
 \begin{equation}
\label{eq17}
P_\nu(amb)=Gk(T_{\rm RX}+T_{\rm amb}).
\end{equation}
The antenna temperature as calibrated by the chopper wheel method $T_{\rm C}$ is given by
$$T_{\rm C}=\frac{P_\nu(source)-P_\nu(sky)}{P_\nu(amb)-P_\nu(sky)}T_{\rm amb}$$
\begin{equation}
\label{eq18}
~~~~~~~~~~~~~~~=\frac{T_{\rm A} T_{\rm amb}}{(1-\zeta) T_{\rm amb}-\eta(1-e^{-\tau})T_{\rm atm}}.
\end{equation}
If we assume $T_{\rm amb}=T_{\rm atm}$ and $\kappa=0$, the antenna temperature is equal to the corrected antenna temperature, $T_{\rm C}=e^{\tau}T_{\rm A}/\eta=T_{\rm A}^\ast$. However we assume here only that $T_{\rm amb}=T_{\rm atm}$. The measured antenna temperature is given by
\begin{equation}
\label{eq19}
T_{\rm C}=\frac{T_{\rm A}}{1-\zeta-\eta+\eta e^{-\tau}}=\frac{T_{\rm A}}{\kappa+\eta e^{-\tau}}.
\end{equation}
The feed efficiency of the 45-m telescope is $\eta\sim0.9$ (\cite{Ukita1994}). The optical thickness at 43~GHz was $\tau\lesssim0.1$. In accordance with the above discussion, the observed decrease in antenna temperature by the metal mesh can be interpreted as an effect of $\kappa$, which is estimated to be as small as $\kappa \lesssim 0.05$. 

Next, we assume $T_{\rm amb}=T_{\rm atm}$ and $\kappa=0$. The system temperature is given by 
\begin{equation}
\label{eq20}
 T_{\rm sys}=\frac{P_\nu(sky)e^{\tau}}{Gk\eta}=\frac{(T_{\rm RX}+T_{\rm amb})e^{\tau}}{\eta}-T_{\rm amb}.
 \end{equation}
Assuming constant $\tau$, $T_{\rm RX}$, and $T_{\rm amb}$ during the observation, the increase in system noise temperature is described as 
\begin{equation}
\label{eq21}
\Delta T_{\rm sys}=e^{\tau} (T_{\rm RX}+T_{\rm amb})\left[\frac{1}{\eta+\Delta\eta}-\frac{1}{\eta}\right].
 \end{equation}
The observed increase of the system temperature corresponds to the 12\% decrease of the efficiency, $\Delta\eta=-0.12$. This is consistent with the measured reflection coefficient described in the previous section. 

\section{Summary}\label{sec:summary}
We carried out precise measurements of millimeter-wave (Q band) RF performance for metal mesh materials in space radio telescope antennas. The results are summarized as follows:

(1) The reflection coefficient and the radiative coefficient significantly vary depending on the tension, side, and rotation angle of the metal mesh, whereas the transmission coefficient is independent of these conditions and has almost constant values.

(2) The increase of the reflection coefficient and the decrease of the radiative coefficient as a function of the tension could be explained by the lower contact resistance between wires of the metal mesh due to the tighter tension. 
Dependence of the rotation angle could be interpreted as a result of anisotropy of the mesh knitting pattern and overlapping pattern of the wires. 

(3) We have estimated deterioration of the $T_{\rm sys}$ caused by losses of the radiative coefficient and the transmission coefficient. The minimum estimated increase of the $T_{\rm sys}$ is as high as 22~K, which makes the total $T_{\rm sys}$ considerably larger than the requirement for the ASTRO-G LDR. Therefore, not only the tension and rotation angle of the metal mesh surface but also its mechanical structure should be carefully considered when using a metal mesh antenna surface for space radio telescopes in the Q band or higher frequencies.  

(4) Astronomical observations at the Nobeyama 45 m radio telescope toward a SiO maser source with metal mesh material in the signal path indicate decreased peak intensity and increased $T_{\rm sys}$ compared with the case without the metal mesh. These results can be interpreted as an effect of scattered radiation terminated by the sky and of degradation in feed efficiency by the metal mesh, respectively.


\bigskip 
The authors thank ASTRO-G project manager Prof. H. Saito, and Prof. K. Higuchi at ISAS/JAXA, for their continuous encouragement. They also thank Dr. A. Doi at JAXA/ISAS, Dr. A. Miyazaki at the National Astronomical Observatory of Japan (NAOJ), Mr. W. Endo and Ms. M. T. Sato at the University of Tokyo, and Mr. M. Kimura at Hosei University for assisting in measurements and observations, and for helpful discussions.

\end{document}